\pdfoutput=1

\documentclass[aps,twocolumn,amsmath,amssymb,preprintnumbers,superscriptaddress,floatfix]{revtex4}
\usepackage[utf8]{inputenc}
\usepackage{bm}
\usepackage{siunitx}

\usepackage{enumerate}
\usepackage{amsfonts}
\usepackage{amsmath}
\usepackage{amssymb}
\usepackage{color}
\usepackage{soul}

\usepackage{graphicx}

\usepackage[colorlinks,allcolors=blue]{hyperref}
\usepackage[capitalize]{cleveref}

\let\phi=\varphi
\let\epsilon=\varepsilon

\newcommand{\fermi}{\text{F}}

\definecolor{DarkRed}{rgb}{0.80,0,0}
\definecolor{DarkGray}{rgb}{0.7,0.7,0.7}

\newcommand{\prlsection}[1]{\textit{#1}.\kern0.05em---\kern0.05em\ignorespaces}

\begin{document}
\title{Controlling Majorana modes by $p$-wave pairing in two-dimensional $p+id$ topological superconductors}
\author{Morten Amundsen}
\affiliation{Nordita, KTH Royal Institute of Technology and Stockholm University,
Hannes Alfvéns väg 12, SE-106 91 Stockholm, Sweden}
\author{Vladimir Juri\v{c}i\'c}
\affiliation{Nordita, KTH Royal Institute of Technology and Stockholm University,
Hannes Alfvéns väg 12, SE-106 91 Stockholm, Sweden}
\affiliation{Departamento de F\'isica, Universidad T\'ecnica Federico Santa Mar\'ia, Casilla 110, Valpara\'iso, Chile}
\begin{abstract}
We show that corner Majorana zero modes in a two-dimensional $p+id$ topological superconductor  can be controlled by the manipulation of the parent $p-$wave superconducting order.
Assuming that the $p$-wave superconducting order is in either a chiral or helical phase, we find that when a  $d_{x^2-y^2}$ wave superconducting order is induced, the system exhibits  quite different behavior depending on the nature of the parent $p$-wave phase. In particular, we find that while in the helical phase, a localized Majorana mode appears at each of the four corners, in the chiral phase, it is localized only along two of the four edges. We furthermore demonstrate that the Majoranas can be directly controlled by the form of the edges, as we explicitly show in case of the circular edges.
We argue that the application of strain may provide additional means of fine-tuning the Majorana zero modes in the system, in particular, it can partially gap them out. Our findings may be relevant for probing the  topology in two-dimensional mixed-pairing superconductors.
\end{abstract}

\maketitle


\section{Introduction}
Majorana zero modes (MZMs) represent a hallmark feature of topological superconductivity with several interesting properties~\cite{Read2000,Kitaev2001,Hasan2010,Qi2011,Tanaka2012,Leijnse2012,Sato2017,LinderRMP2019}.
In addition to their fundamental importance, MZMs are fascinating because they exhibit non-Abelian statistics, which manifests in braiding operations. This can be of crucial importance for future applications, for instance in information technology~\cite{Kitaev2003,Nayak2008,Aguado2020,Oreg2020}, and has been the subject of intense research~\cite{Alicea2012,Beenakker2013,Vijay2016,Karzig2017,Plugge2017,Lutchyn2018,Haim2019,Park2020,Prada2020,Huang2021,Flensberg2021}. The MZMs appear as topologically protected boundary  states at interfaces where the topological invariant changes, while the bulk remains gapped. Typically, a topological superconductor (SC) features MZMs at interfaces of codimension $m=1$, imprinting the standard topological bulk-boundary correspondence~\cite{Hasan2010,Qi2011}. In a one-dimensional finite system, they take the form of localized modes at the ends, while in two and three dimensions they are realized as, respectively, edge and surface states.

New platforms for the realization of the MZMs have recently appeared with the advent of higher-order topological states~\cite{Benalcazar2017,Benalcazar-prb2017,Song2017,Schindler2018,Trifunovic2017,Gong2018,Khalaf2018,Calugaru2019,Fulga2019,Agarwala2020,Szabo2020} which generalize the standard bulk-boundary correspondence. Within this class of states, higher-order topological SCs can localize MZMs at interfaces of codimension $m > 1$. In particular, two-(three-)dimensional second-(third-)order topological SC may host corner MZMs~\cite{Wang2018,Wu-yan2019,Wang-liu2018,Liu2018,Yan2018,Volpez2019,Yan2019,Zhu2019,Pan-yang-chen-xu-liu-liu-HOTSC,Ghorashi-HOTSC,Broyrantiunitary,Trauzettel-HOTSC,Bjyang-HOTSC-1,Bomantara-HOTSC,BroysoloHOTSC2020,SBZhang2020,Sigrist2020,Thomale2020PRX,Tiwari2020,Ghoshnagsaha2021,Shen2021,Hsu2020,Hsu2018,YanPRB2019,Jack2019,Wu2020,Kheirkhah2020,Manna2020,Zhu2018,Sigrist2020,Ikegaya2020,Franca2019,Kheirkhah2021,Ghosh2021,RoyODI2021} with codimension $m=d$ in  $d$ spatial  dimensions.  In a two-dimensional (2D) topological  state with an insulating or superconducting bulk gap, the corner zero modes  can be obtained  by gapping out the first-order edge states with a mass term that features a domain wall in momentum space, realizing a special case of the hierarchy of higher-order topological states ~\cite{Calugaru2019,Nag2021}. In this respect, several concrete ways for the realization of  corner MZMs in two dimensions  have been proposed so far, as, for instance,
by inducing  a superconducting gap for the edge states of a topological insulator, either intrinsically~\cite{Hsu2020}, or via the proximity effect~\cite{Yan2018, Wang2018, Hsu2018,Yan2019,Liu2018,Jack2019,Wu2020}. It has also been shown that they may emerge when pairing of an appropriate symmetry is combined with a spin-dependent field, such as spin-orbit coupling~\cite{Zhu2019, Kheirkhah2020, Manna2020}. Furthermore, several works have recently proposed means by which the order of a topological superconductor may be manipulated, along with the position of the resulting corner MZMs. Indeed, a first-order topological SC may be promoted to the second order by the application of a magnetic field, with the location of the corner modes determined by the orientation of the field~\cite{Zhu2018,Sigrist2020,Ikegaya2020}. It has also been theoretically shown that second-order topological superconductivity can emerge in Josephson junctions, in which case the phase difference between the SCs provides additional means of manipulation~\cite{Volpez2019,Franca2019}.

We here consider a different scenario in which a variety of MZMs can be generated solely by manipulating the parent $p$-wave superconducting order in a mixed parity $p+id$ 2D SC. We assume that the $p$-wave superconducting order, in the absence of any other pairing, hosts a first-order topological state, and exists in either a chiral or helical phase, referring to whether it breaks or preserves time reversal symmetry, respectively.  We show that when a  $d_{x^2-y^2}$ wave superconducting order is induced e.g. via the proximity effect, the system exhibits  quite different behavior depending on the parent $p$-wave phase. In particular, we find that in the helical phase, a localized Majorana mode appears at each of the four corners, as shown in Fig.~\ref{fig:helldos}. In the chiral phase, on the other hand,  no corner modes appear. Instead, a gap emerges in two out of the four edge modes (Fig.~\ref{fig:childos}). Therefore, the behavior of the Majorana modes can be tuned solely by manipulating  the pairing symmetry of the parent topological SC. As we show, the edge geometry can also be relevant in this regard (see Fig.~\ref{fig:childoscirc}), where we display the Majorana states for a circular edge geometry. Finally, we demonstrate that the application of strain may drive a topological phase transition when the parent phase is chiral. As a result, the strain gaps out two out of the four edges, as displayed in Fig.~\ref{fig:strain}.

The rest of the paper is organized as follows. In \cref{sec:model} we introduce the model for the $p+id$ topological SC we consider. Next, we present analytical arguments for the behavior of the resulting edge states in \cref{sec:analytical}, before moving on to discuss numerical results in \cref{sec:results}. Finally, we analyze the effect of strain in \cref{sec:strain} and present our conclusions together with an outlook in \cref{sec:conclusions}.

\section{Model} \label{sec:model}
We employ the standard Bogoliubov--de~Gennes formalism to study the system, with the Hamiltonian given as
\begin{align}\label{eq:Hamiltonian1}
H = \frac{1}{2}\sum_{\bf k} \psi_{\bf k}^\dagger\hat{H}({\bf k})\psi_{\bf k},
\end{align}
where the corresponding Nambu spinor is   $\psi_{\bf k} = \begin{pmatrix} c_{{\bf k}\uparrow} & c_{{\bf k}\downarrow} & c_{-{\bf k}\uparrow}^\dagger & c_{-{\bf k}\downarrow}^\dagger\end{pmatrix}^\top$, with $c_{{\bf k}\uparrow}$ ($c_{{\bf k}\downarrow}^\dagger$) as the annihilation (creation) operator for the quasiparticle with spin up (down) and momentum ${\bf k}$ . Here,
\begin{align}
\hat{H} = \begin{pmatrix} h(\bm{k}) & \Delta(\bm{k}) \\ -\Delta^*(-\bm{k}) & -h^*(-\bm{k}) \end{pmatrix},
\label{eq:h1}
\end{align}
with the blocks describing the normal (non-superconducting) state and the pairing, respectively, given by
\begin{align}
h(\bm{k}) =& \left(\frac{\hbar^2\bm{k}^2}{2m} - \mu\right)\sigma_0\equiv \xi_k \sigma_0, \label{eq:h2}\\
\Delta(\bm{k}) =&\left[\frac{\Delta_p}{k_{\fermi}} \bm{g}(\bm{k})\cdot\bm{\sigma} + \frac{i\Delta_d}{k^2_{\fermi}}\left(k_x^2 - k_y^2\right)\right]i\sigma_2,\label{eq:h3}
\end{align}
where $\mu$ is the chemical potential, $m$ is the quasiparticle mass, the Pauli matrices ${\bm \sigma}$ and the unit $2\times2$ matrix, $\sigma_0$, act in the spin space, $k_F$ is the Fermi momentum. Here,  $\Delta_p$ and $\Delta_d$ are the amplitudes of the $p$- and $d$-wave superconducting orders. For the latter we choose $d_{x^2-y^2}$ component $\sim(k_x^2-k_y^2)$ as it features domain walls along the diagonals in the momentum space, located at $k_x=\pm k_y$, where it changes the sign, and thereby \emph{partially} gapping out the edge states~\cite{footnote}. We also include a relative phase of $\pi/2$ between the two order parameters, implying that the mixed pairing state breaks time-reversal symmetry. Notice, however, that the $d-$wave component preserves  the product of the $C_4$ rotational  and time-reversal ($\mathcal{T}$) symmetries, $C_{4T}=C_4\mathcal{T}$ implying that the resulting $p+id$ superconducting state may feature the same composite symmetry. This, indeed, occurs when the $p-$wave component  is helical (see below), in which case also a $Z_2$ topological invariant protects the resulting second-order topological SC~\cite{Wang2018}. The vector $\bm{g}(\bm{k})$ parametrizes  the triplet $p-$wave superconducting paring, and takes the form
\begin{align}
\bm{g}(\bm{k}) = \cos\theta\;\bm{k}\times\hat{z} + \sin\theta \left(k_x + ik_y\right)\hat{z},
\end{align}
where $\hat{z}$ is the unit vector pointing in the $z$ direction, assumed to be normal to the 2D plane. The helical phase is found by setting $\theta = 0$, and represents a time-reversal invariant SC, featuring a pair of counterpropagating gapless Majorana edge modes. On the other hand, the chiral phase, which breaks the time-reversal symmetry, is found for $\theta = \pi/2$. In the following we are interested only in these two special cases, and also refer to the mixed $p + id$ state as either chiral or helical depending upon the phase of the parent $p$-wave component.  Since we only consider the topological regime, we do not include the $s-$wave pairing in the Hamiltonian in Eq.~\eqref{eq:h1}.

The bulk spectrum of the Hamiltonian in \cref{eq:h1,eq:h2,eq:h3},  is given as
\begin{align}
E(\bm{k}) = \pm\sqrt{ \xi_k^2 + a_+\Delta_p^2+a^2_-\Delta_d^2\pm 2\Delta_p\frac{k_y}{k_\fermi}\,b(\theta)},
\label{eq:spectrum}
\end{align}
with $a_\pm\equiv a_\pm(\bm{k}) = \left(k_x^2\pm k_y^2\right)/k_\fermi^2$, and
\begin{align*}
b(\theta) = \sin\theta\sqrt{a_-^2\Delta_d^2 + a_+\Delta_p^2\cos^2\theta}.
\end{align*}
 The spectrum for the helical ($\theta=0$) and the chiral ($\theta=\pi/2$) phases thus  reduces to
\begin{align*}
E(\bm{k}) = \begin{cases}\pm\sqrt{\xi_k^2 + a_+\Delta_p^2 + a_-^2\Delta_d^2}, & \theta = 0 \\
\pm\sqrt{\xi_k^2 + \frac{k^2_x}{k^2_\fermi}\Delta^2_p +  \left(a_-\Delta_d \pm \frac{k_y}{k_\fermi}\Delta_p\right)^2}, & \theta = \frac{\pi}{2}. \end{cases}
\end{align*}
It is clear that for $\theta = 0$, the bulk band structure always features a gap, as long as $\Delta_p\neq 0$. The same holds for $\theta = \frac{\pi}{2}$, except for the critical value of $\Delta_d = \Delta_p$, where the gap closes. We furthermore note that both of the gapped regions $\Delta_d < \Delta_p$ and $\Delta_d > \Delta_p$ are topologically nontrivial and feature gapless edge states, as will be evident from the following.

\section{Analytical results}\label{sec:analytical}
To understand the effects of the $d$-wave superconducting order on the edge states present in this system, we introduce an interface which is oriented with an angle $\alpha$ with respect to the horizontal ($x-$) principal crystalline axis. The Hamiltonian in this rotated coordinate system, $\bm{k}=(k_x,k_y)\to \bm{k}'=(k_{||},k_\perp)$, is given by
\begin{align}
\label{eq:rotH}
\hat{H}'(\bm{k}') = \hat{R}\hat{H}(\mathcal{R}^{-1}\bm{k}')\hat{R}^\dagger,
\end{align}
where
\begin{align}\label{eq:rotation-Nambu}
\hat{R}(\alpha) = \begin{pmatrix} R(\alpha) & 0 \\ 0 & R^*(\alpha) \end{pmatrix},
\end{align}
is the rotation operator in the Nambu basis defined in Eq.~\eqref{eq:Hamiltonian1},
and $R=\exp[i\alpha\sigma_3/2]$ represents the  operator of the rotation by the angle $\alpha$ about the $z$ axis in the spin-$1/2$ representation. The momenta ${{\bm k}'}$, and ${\bm k}$ are related by a rotation about the $z-$axis with the same angle $\alpha$, ${{\bm k}'}= \mathcal{R}(\alpha){\bm k}$, with the  rotational matrix given by
\begin{align}
\mathcal{R}(\alpha) &=\begin{pmatrix} \cos\alpha & \sin\alpha \\ -\sin\alpha & \cos\alpha \end{pmatrix}.
\end{align}
The corresponding edge Hamiltonian is then obtained by projecting a part of the rotated bulk Hamiltonian $\hat{H}'(\bm{k}')$ (Eq.~\ref{eq:rotH}) \emph{not} used to obtain the zero modes  onto the subspace spanned by the zero-energy edge modes.

In the following,  we assume that both $\Delta_p$ and $\Delta_d$ are much smaller than the Fermi energy (weak-pairing limit), which we set equal to the chemical potential, $\mu = \hbar^2k_\fermi^2/2m$. In addition, we assume that the wave vector parallel to the interface is much smaller than the Fermi momentum, $k_{||} \ll k_\fermi$.
To find the form of the edge Hamiltonian, as previously announced, we first perform the rotation in Eq.~\eqref{eq:rotH}   and then solve for the zero-energy modes in the absence of the $d-$wave pairing. To this end, in the rotated Hamiltonian \eqref{eq:h1}, we isolate a part that is proportional to $\Delta_p$ and  depends  on $k_\perp$, the wave vector orthogonal to the interface. The total  Hamiltonian  therefore acquires the form
\begin{align}
\hat{H}'({\bf k}') = \hat{H}'_0(k_{\perp}) + \hat{H}'_1(k_\parallel) + \hat{H}'_2(k_\perp,k_\parallel),
\end{align}
with
\begin{align}
\hat{H}'_0(k_\perp)&\simeq  \frac{\hbar^2}{2m}\left(k_\perp^2 - k_\fermi^2\right)\tau_3\sigma_0\nonumber\\
&-\frac{k_\perp}{k_\fermi}\Delta_p\left(\tau_2\sigma_0\cos\theta - e^{i\alpha\tau_3\sigma_0}\tau_1\sigma_1\sin\theta\right),
\label{eq:unperturbed}\\
\hat{H}'_1(k_\parallel) &= \frac{k_{||}}{k_\fermi}\Delta_p\left(\tau_1\sigma_3\cos\theta - e^{i\alpha\tau_3\sigma_0}\tau_2\sigma_1\sin\theta\right), \\
\hat{H}'_2(k_\perp,k_\parallel) &= -\frac{\Delta_d}{k_F^2}\left[(k_{||}^2-k_\perp^2)\cos2\alpha-2k_{||}k_\perp\sin2\alpha\right]\tau_1\sigma_2.
\label{eq:per2}
\end{align}
Here,  $\tau_a\sigma_b\equiv\tau_a\otimes\sigma_b$ is a Kronecker product between Pauli matrices in Nambu and spin space, respectively. Since the edge breaks translation invariance, we make the substitution $k_\perp \to -i\partial_{r_\perp}$, where $r_\perp$ is the direction orthogonal to the edge, and solve the resulting differential equation with the boundary conditions $\psi(r_\perp = 0) = \psi(r_\perp = \infty) = 0$. The unperturbed Hamiltonian, given by Eq.~\eqref{eq:unperturbed}, admits two degenerate zero-energy eigenstates of the form
\begin{align}\label{eq:zero-modes}
\psi_a(r_\perp) = \sqrt{2\kappa}\,\,e^{-\kappa r_\perp}\sin k_\fermi r_\perp \,\,\phi_a,
\end{align}
with $a=1,2$, the spinors
\begin{align*}
\phi_1 =& \begin{pmatrix} -i\cos\theta & e^{i\alpha}\sin\theta & -i & 0\end{pmatrix}^\top, \\
\phi_2 =& \begin{pmatrix} e^{i\alpha}\sin\theta & -i\cos\theta & 0 & -i\end{pmatrix}^\top,
\end{align*}
and $\kappa=m\Delta_p/(\hbar^2k_F)$.
The effective Hamiltonian for the edge states is thus found, to first order in the perturbation expansion as the projection
\begin{align}\label{eq:projection-edge}
H^e_{ab}(\theta) = \int_{0}^\infty dr_\perp\; \psi_a^\dagger(r_\perp)(\hat{H}'_1+\hat{H}'_2)\psi_b(r_\perp),
\end{align}
which gives

\begin{align}
&H^e(\theta) = -\Delta_p\frac{k_{||}}{k_\fermi}\begin{pmatrix}1 & -\frac{i}{2}e^{i\alpha}\sin2\theta \\ \frac{i}{2}e^{-i\alpha}\sin2\theta & -\cos 2\theta\end{pmatrix} \nonumber\\
&-\Delta_d\cos2\alpha \begin{pmatrix}\sin\theta\cos\alpha & i\cos\theta \\ -i\cos\theta & -\sin\theta\cos\alpha\end{pmatrix},
\label{eq:hedge}
\end{align}
where we have kept only the leading term in  \cref{eq:per2}.  The term $\sim k_\parallel$  in  \cref{eq:per2}  vanishes identically after taking  the projection \eqref{eq:projection-edge},     and we have neglected the term $\sim k_\parallel^2$ as being less relevant by power counting than the leading one in the low-energy edge Hamiltonian. Note that this term can be,  in principle, included in this Hamiltonian following the outlined procedure.

\begin{figure}[t]
\includegraphics[width=\columnwidth]{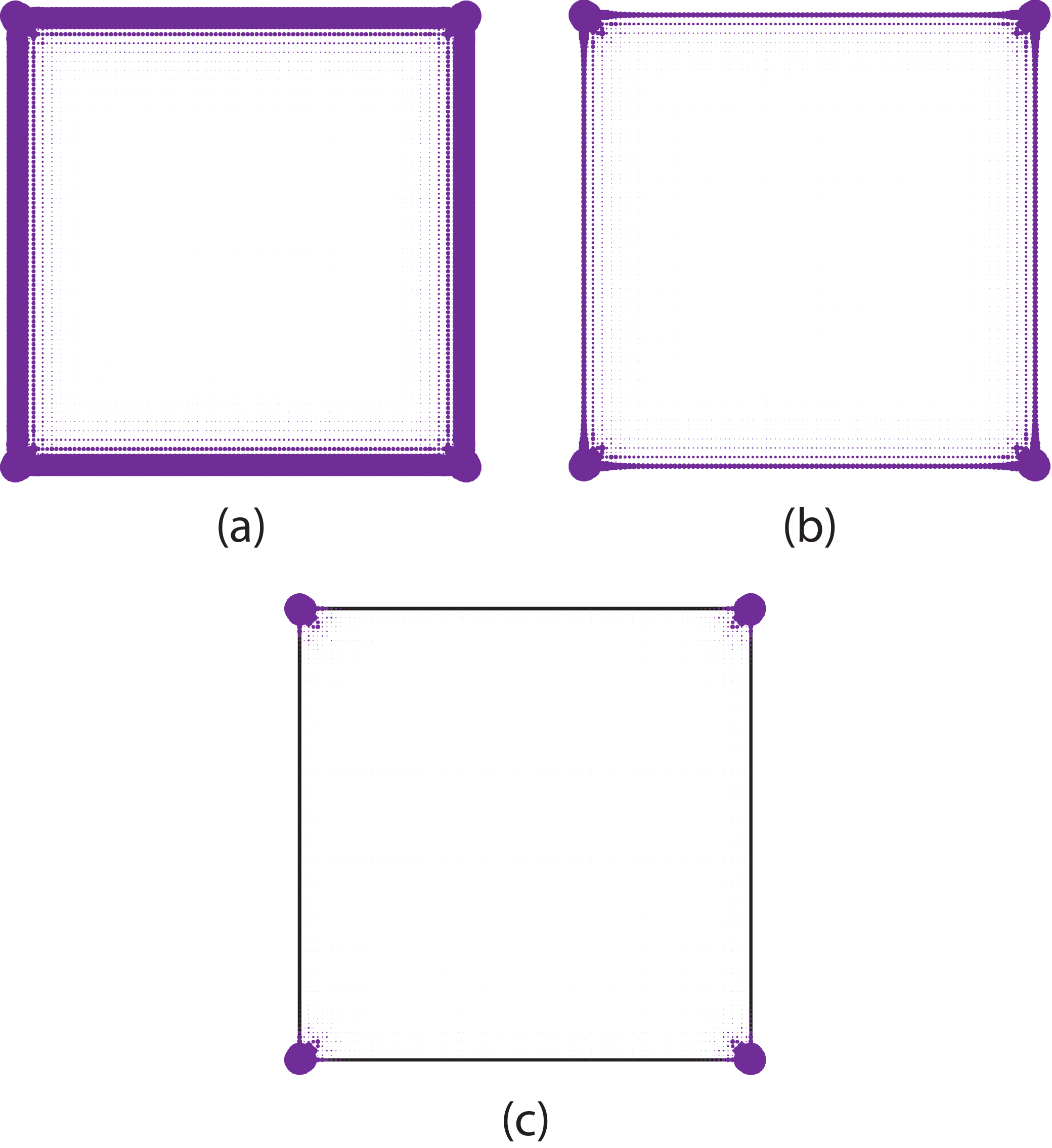}
\caption{Local density of states at zero energy, $\nu(0)$, in the helical phase ($\theta = 0$), with the $d$ wave order parameter set to (a) $\Delta_d = 0$, (b) $\Delta_d = 0.1\Delta_p$, and (c) $\Delta_d = 0.2\Delta_p$. The dot size at a particular point indicates the size of $\nu(0)$ at that point. In all plots, we use the Hamiltonian in Eq.~\eqref{eq:HamLattice} and   set $\mu = 2t$ and $\Delta_p = t=1$, and $a=1$. The system size is $100\times 100$ sites.  }
\label{fig:helldos}
\end{figure}

The first term in \cref{eq:hedge} describes the Majorana edge modes, having a characteristic nodal structure in $k_{\parallel}$ in the absence of the $d-$wave pairing. These modes are, in turn,  gapped by the $d-$wave pairing term. For $\theta = 0$ (the helical phase), one thus obtains
\begin{align}
H^e(\theta = 0) = \begin{pmatrix} -\Delta_p\frac{k_{||}}{k_\fermi} & i\Delta_d\cos2\alpha \\ -i\Delta_d\cos2\alpha & \Delta_p\frac{k_{||}}{k_\fermi} \end{pmatrix}.
\label{eq:helicaledge}
\end{align}
Clearly, a mass term proportional to $\cos2\alpha$ appears, having domain walls along the two diagonals, located at $\alpha=\pm\pi/4$. Hence, the edge modes are gapped out, and only the corner modes remain, in agreement with Refs.~\cite{Wang2018,Wu-yan2019}. These corner MZMs are protected by the composite $C_{4T}$ symmetry and a $Z_2$ topological invariant [see also the discussion after Eq.~\eqref{eq:h3}].

In the chiral phase, for $\theta = \pi/2$, we do not have a band crossing which may turn into an anti-crossing and open up a gap at the edge. In this case \cref{eq:hedge} takes the form
\begin{align}
H^e\left(\theta = \frac{\pi}{2}\right) = -\Delta_p\frac{k_{||}}{k_F}\sigma_0-\Delta_d \cos\alpha \cos 2\alpha\,\, \sigma_3,
\label{eq:chiraledge}
\end{align}
which suggests that the presence of the $d$ wave order parameter only has a trivial effect on the edge states in the regime $\Delta_d<\Delta_p$. However, the closing of the bulk gap at $\Delta_d = \Delta_p$ signals a possible phase transition and therefore something interesting may occur also in this system. It turns out that a \emph{selective gapping} of the edge states is possible for $\Delta_d > \Delta_p$, as can already be seen from \cref{eq:chiraledge}. Namely,  by taking the values of the angle $\alpha=0$ ($\alpha=\pi/2$), corresponding to horizontal (vertical) edge, we obtain that the horizontal and vertical edge should be gapped and gapless, respectively.  This is exactly what we find by numerical means, as shown  in the next section.

\section{Numerical Analysis} \label{sec:results}

In addition to the  analytical analysis presented in the previous section, we also numerically study the $p+id$ SC described by \cref{eq:h1,eq:h2,eq:h3}. The corresponding square lattice   Hamiltonian reads
\begin{align}\label{eq:HamLattice}
H = \frac{1}{2}\sum_{jl}\psi^\dagger_j \hat{H}_{jl}\psi_l,
\end{align}
with Nambu vector at the lattice site $j$,  $\psi_j = \begin{pmatrix} c_{j\uparrow} & c_{j\downarrow} & c^\dagger_{j\uparrow} & c^\dagger_{j\downarrow}\end{pmatrix}^{\top}$, and $\hat{H} = \hat{H}_0 + \hat{H}_p + \hat{H}_d$, where
\begin{align}
\hat{H}_0 =& \left[-t\left(\delta_{\hat{x}} + \delta_{-\hat{x}}+\delta_{\hat{y}} + \delta_{-\hat{y}} \right) +\left(4t - \mu\right)\delta_{jl}\right]\tau_3\sigma_0 \\
\hat{H}_p =& \frac{\Delta_p}{2ik_\fermi a}\left[\left(\vphantom{\delta^\dagger}\left(\delta_{\hat{y}} - \delta_{-\hat{y}}\right)\tau_1\sigma_3-\left(\delta_{\hat{x}} - \delta_{-\hat{x}}\right)\tau_2\sigma_0\right)\cos\theta\right. \nonumber \\
&\left.+\left(\vphantom{\delta^\dagger}\left(\delta_{\hat{x}} - \delta_{-\hat{x}}\right)\tau_1\sigma_1-\left(\delta_{\hat{y}} - \delta_{-\hat{y}}\right)\tau_2\sigma_1\right)\sin\theta\right] \\
\hat{H}_d =& \frac{\Delta_d}{k_\fermi^2 a^2}\left[\delta_{\hat{x}} + \delta_{-\hat{x}} - \delta_{\hat{y}} - \delta_{-\hat{y}}\right]\tau_1\sigma_2.
\end{align}
In the above, $t = \hbar^2/2ma^2$, and we use the shorthand notation $\delta_{\hat{n}} \equiv \delta_{j+\hat{n},l}$. We furthermore remark that the momentum space representation of the above lattice Hamiltonian with periodic boundary conditions is found from \cref{eq:h1,eq:h2,eq:h3} by replacing
\begin{align*}
&\xi_k \to -2t\left(\cos k_xa + \cos k_ya\right) +4t-\mu,\\
&\bm{k}\to \frac{1}{a}\sin(\bm{k}a),
\end{align*}
 and substituting
 \begin{equation*}
 k_x^2-k_y^2\to \frac{2}{a^2}\left(\cos k_xa - \cos k_ya\right)
 \end{equation*}
 in the $d$-wave component the superconducting order parameter. This has the effect of shifting the location at which the bulk gap closes, which is a characteristic feature of the chiral phase, to
\begin{align}
\Delta_d = \Delta_p\sqrt{1 - \frac{\mu}{4t}} \equiv \Delta_d^c.
\end{align}
Furthermore, with $\mu/t = k_\fermi^2a^2$, it is manifest that the above discretized model becomes equivalent to its continuum counterpart in the limit $k_\fermi a \ll 1$.

\begin{figure}[t]
\includegraphics[width=\columnwidth]{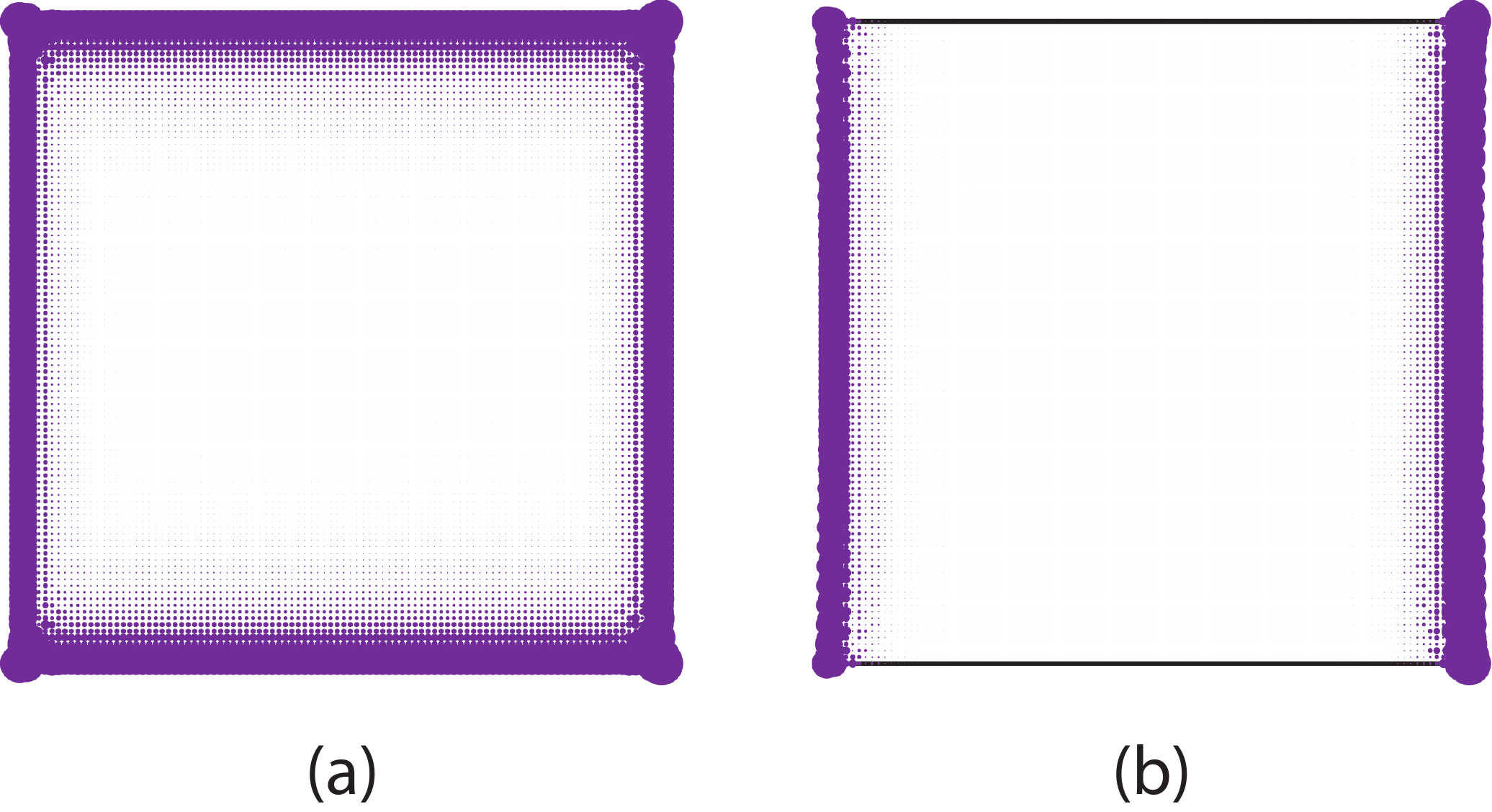}
\caption{The local density of states in the chiral phase $(\theta = \frac{\pi}{2})$ in the two regimes separated by  the critical value of the $d$ wave order parameter $\Delta_d^c$, at which the bulk gap closes. In (a) $\Delta_d = 0.5\Delta_d^c$, whereas in (b) $\Delta_d = 1.5\Delta_d^c$. The dot size at a particular point indicates the size of $\nu(0)$ at that point. In both plots, we use the Hamiltonian in Eq.~\eqref{eq:HamLattice} and   set $\mu = 2t$ and $\Delta_p = t=1$, and $a=1$. The system size is $100\times 100$ sites.}
\label{fig:childos}
\end{figure}

We now consider the edge states in a square-lattice system. We compute the local density of states at site position $j$ as
\begin{align}
\nu_j(E) &= \sum_n |v_{n,j}|^2\delta(E- E_n) \nonumber \\
&\simeq \frac{1}{\sqrt{\pi\lambda}}\sum_n |v_{n,j}|^2 e^{-\left(E - E_n\right)^2/\lambda},
\end{align}
where $E_n$ are the eigenvalues of the Hamiltonian, and $v_{n,j}$ the values of the corresponding eigenvectors at $j$. The broadening parameter $\lambda$ is set to \num{5d-3}. In  the helical phase,   as shown in \cref{fig:helldos},  the edge states quickly vanish with increasing $\Delta_d$, being completely gapped out at $\Delta_d = 0.2\Delta_p$, and thus leaving only the corner modes.

\begin{figure}[t]
\includegraphics[width=\columnwidth]{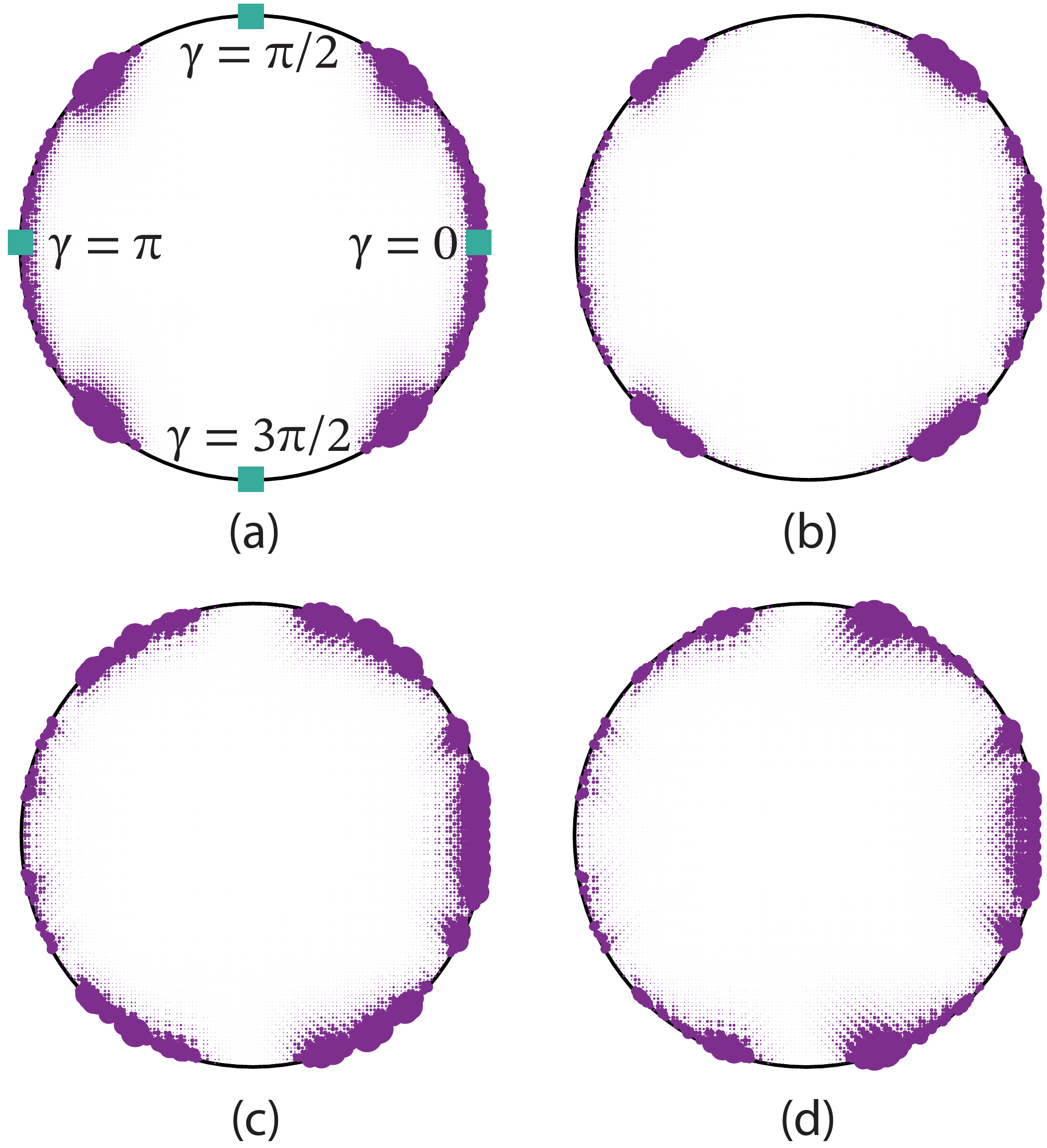}
\caption{The edge states in a circular geometry when the system is in the chiral phase for increasing values of the $d$ wave order parameter, equal to (a) $\Delta_d = 1.5\Delta_d^c$, (b) $\Delta_d = 3\Delta_d^c$, (c) $\Delta_d = 4.5\Delta_d^c$, and (d) $\Delta_d = 6\Delta_d^c$. The square markers in panel (a) indicates the opening angles probed by the square geometry shown in \cref{fig:childos}. The dot size at a particular point indicates the size of $\nu(0)$ at that point. In all plots we have set $\mu = 2t$, $\Delta_p = t=1$, $a=1$ in the Hamiltonian in Eq.~\eqref{eq:HamLattice}. The system consists of a disk with a radius of 60 sites.  }
\label{fig:childoscirc}
\end{figure}

We turn to the chiral phase, where the gap closing at $\Delta_d = \Delta_d^c$ separates  two regions of interest. The region with $\Delta_d < \Delta_d^c$ is topologically equivalent to the case where $\Delta_d = 0$, and we thus expect that the results from \cref{eq:hedge} apply here, implying that the $d$-wave order parameter does not gap out the chiral edge states. This is indeed found to be the case in our numerical analysis, as illustrated in \cref{fig:childos}(a), in which the local density of states at $\Delta_d = t/2\sqrt{2} = 0.5\Delta_d^c$ is shown. In contrast, for $\Delta_d = 3t/2\sqrt{2} = 1.5\Delta_d^c$, shown in \cref{fig:childos}(b), the behavior is different. In that case the horizontal edges are gapped out, but the vertical edge states remain gapless, in agreement with Eq.~\eqref{eq:chiraledge}. We furthermore note that a phase shift of ${\pi}/{2}$ in the relative phase between $\Delta_d$ and $\Delta_p$ would amount to a $\pi/2$ rotation of the result in \cref{fig:childos}(b). Therefore, the relative phase between the two pairing order parameters translates into the pattern of the gap at the edge of the system.

\begin{figure}[t]
\includegraphics[width=\columnwidth]{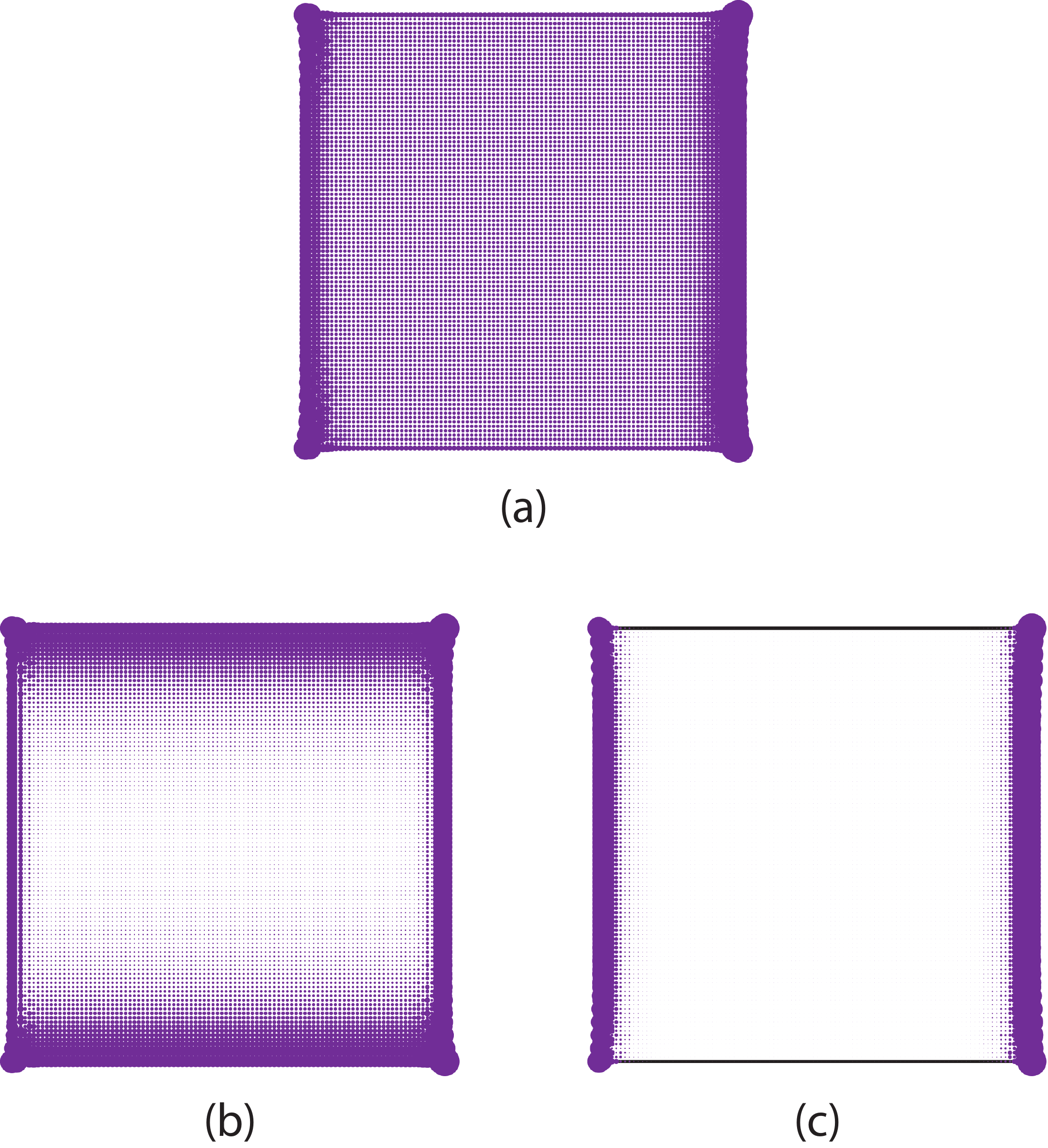}
\caption{The effect of strain on the zero-energy edge states in the chiral phase. In (a) the local density of states at zero energy is shown at $\Delta_d = \Delta_d^c$ without any strain. In (b) and (c) the same system is shown for an applied uniaxial strain of $\epsilon_{xx} = \SI{10}{\percent}$, and $\epsilon_{yy} = \SI{10}{\percent}$, respectively. The dot size at a particular point indicates the size of $\nu(0)$ at that point. In all plots, we use the  we have set $\mu = 2t$, $\Delta_p=t=1$, and $a=1$ in the Hamiltonian in  Eq.~\eqref{eq:HamLattice}. The system size is $100\times 100$ sites.}
\label{fig:strain}
\end{figure}

We now investigate the edge states in the case of a  disk geometry. This is relevant because the behavior of the edge states for any polygonal geometry may be immediately deduced by comparing the corner opening angles with corresponding points on the circle.  The modeling of the disk is performed by creating a square grid, and discarding all nodes which fall outside a selected radius, here chosen to be 60 nodes. The results are shown in \cref{fig:childoscirc} for increasing values of $\Delta_d$ above the critical value, $\Delta_d^c$. Below $\Delta_d^c$ (not shown), the edge states are uniformly distributed around the entire edge, as expected. Immediately after crossing the critical value, a gap is opened up in the edge states at angles around $\gamma = \left\{\pi/2,3\pi/2\right\}$,   as shown in \cref{fig:childoscirc}(a) for $\Delta_d = 1.5\Delta_d^c$, and $\gamma$ is the polar angle of the circle. This is consistent with the partial gapping of the edge states observed in \cref{fig:childos}(b), which probes the same angles, along with the angles $\left\{0,\pi\right\}$, which are gapless. A further increase in $\Delta_d$ increases the modulation of the density of states along the edge, and produces additional gapped regions, as can be seen in \cref{fig:childoscirc}(b)-(d), which correspond to $\Delta_d/\Delta_d^c = 3$, 4.5, and 6, respectively. Furthermore, the gapped circle sector surrounding $\gamma = \{\pi/2,{3\pi}/{2}\}$ is seen to narrow as $\Delta_d$ becomes larger, but never closes completely, consistent with the domain wall structure of the $d_{x^2-y^2}$-wave pairing.

\section{The effect of strain}
\label{sec:strain}
We investigate strain as a potential means to manipulate the edge states.  We model its effects by introducing a small strain field to the system,
\begin{align}
\epsilon = \begin{pmatrix} \epsilon_{xx} & \epsilon_{xy} \\ \epsilon_{yx} & \epsilon_{yy}\end{pmatrix}.
\end{align}
Here, $\epsilon_{xx}$ and $\epsilon_{yy}$ represent axial strain, defined as positive for tensile strain, and $\epsilon_{xy} = \epsilon_{yx}$ represents shear strain. In the presence of such a strain field, the spatial coordinates transform as $r_i' = \left(\delta_{ij} + \epsilon_{ij}\right)r_j$, which implies that, to linear order in the strain tensor, the momentum transforms as
\begin{align}
k_i' = \left(\delta_{ij} - \varepsilon_{ij}\right)k_j.
\label{eq:kstrain}
\end{align}
By replacing $\bm{k}\to\bm{k}'$ in \cref{eq:h1,eq:h2,eq:h3}, then inserting \cref{eq:kstrain} and retaining only terms up to first order in $\epsilon$, we find that additional  strain-dependent terms are introduced in the Hamiltonian, which may have an effect on the edge states. For an edge with an arbitrary orientation $\alpha$, the corresponding Hamiltonian, after incorporating the effects of strain, read
\begin{align}
&H^e = -\Delta_p\Gamma_p(\epsilon,\alpha)
\frac{k_{||}}{k_\fermi}\begin{pmatrix}1 & -\frac{i}{2}e^{i\alpha}\sin2\theta \\ \frac{i}{2}e^{-i\alpha}\sin2\theta & -\cos 2\theta\end{pmatrix} \nonumber\\
&-\Delta_d\Gamma_d(\epsilon,\alpha) \begin{pmatrix}\sin\theta\cos\alpha & i\cos\theta \\ -i\cos\theta & -\sin\theta\cos\alpha\end{pmatrix},
\label{eq:hstrainedge}
\end{align}
with
\begin{align}
\Gamma_p(\epsilon,\alpha) &=1 - \bar{\epsilon} + \frac{1}{2}\delta\epsilon\cos 2\alpha +\epsilon_{xy}\sin 2\alpha\\
\Gamma_d(\epsilon,\alpha) &= \left(1-2\bar{\epsilon}\right)\cos 2\alpha - \delta\epsilon,
\end{align}
where $\bar{\epsilon} = (\epsilon_{xx} + \epsilon_{yy})/2$ and $\delta\epsilon = \epsilon_{xx} - \epsilon_{yy}$.

We see that strain produces an effective renormalization of the $p$ and $d$ wave order parameters in the edge Hamiltonian, by $\Gamma_p$ and $\Gamma_d$ respectively. Furthermore, for anisotropic strain, $\Gamma_d$ acquires a contribution independent of $\alpha$. This implies that the corner modes in the helical phase, given by \cref{eq:helicaledge} can be gapped out by strain. This is also consistent with the breaking of the $C_{4T}$ symmetry by strain, which protects the topological state. However, the gapping only occurs once the mass domain wall at the corners is removed, which requires rather large strains. For instance, with a uniaxial tensile strain of $\epsilon_{xx} = \epsilon$, the corner modes are gapped out for $\epsilon > \SI{50}{\percent}$. This is much larger than the capacity of any known material, with graphene as the closest contender, reported  to sustain strains of up to $\SI{25}{\percent}$~\cite{Lee2008}. In any case, these levels of strain are certainly far beyond the linear strain regime considered herein, implying the stability of the corner modes  against strains in the experimentally realizable range.

In the chiral phase, we consider  the effect of strain by numerical means. To this end, we replace the lattice parameter $a$ with directionally dependent equivalents, $a_x$ and $a_y$, satisfying $a_i = (1 + \epsilon_{ii})a$. We ignore shear strain. In the regime $\Delta_d > \Delta_d^c$, we can conclude  from \cref{eq:hstrainedge} that $\Delta_d$ and $\Delta_p$ are renormalized slightly differently by strain, which agrees with the numerical analyses for $\Delta_d < \Delta_d^c$. Hence, strain may be used to change the ratio between the amplitudes of the $d$ and $p$ wave superconducting order, and thus cause a transition between the two phases exhibiting a different behavior of the  edge states, which agrees with our numerical analysis.  Setting $\Delta_d = \Delta_d^c$, which in the unstrained case is gapless, as shown in \cref{fig:strain}(a), we perturb the system by applying a uniaxial strain of \SI{10}{\percent} along the $x$ and $y$ directions, respectively shown in \cref{fig:strain}(b) and (c). In the former case, it can be seen that the system is pushed into the state with uniform gapless edge modes, whereas in the latter case, the selectively gapped state is entered. Therefore, in the chiral phase, strain can be used to tune the form and the localization of the edge states.

\section{Conclusions and Outlook}
\label{sec:conclusions}

In conclusion, we have studied the Majorana zero modes in a 2D SC with mixed $p$- and $d$-wave pairings, considering  both  the helical and chiral $p$-wave phases. For the parent  helical $p$-wave order, we found that the effect of the  $d$-wave order parameter is to  gap out the edge states, leaving only zero-energy Majorana corner modes. In the case of the parent chiral $p$-wave pairing, we showed that the edge states can be \emph{partially} gapped out above a critical value of the $d$-wave order parameter. Therefore, the parent $p$-wave phase can control the form of the Majorana modes in the $p+id$ topological SC.  We found that the localization of the Majorana modes can also be tuned by the geometry of the edges, as implied  by our results in the circular edge geometry. Moreover,  the localization of the MZMs in a polygonal geometry may  be  inferred  by  identifying  the  opening  angle  of a given corner with a corresponding point on the circle.  We have also investigated the effect of strain and found that the higher-order topological SC produced in the parent helical phase is robust against strain up to the experimentally reachable values. On the other hand, in the chiral phase, we showed that strain can be used to induce a transition between topologically distinct phases with gapless and partially gapped edge states.  Hence, the application of strain, through the strain-induced topological phase transition,   may indeed provide a direct means of  manipulating  the Majorana modes.

The experimental realization of such superconducting system is a challenging issue at present, but we will nevertheless indicate a few promising avenues. First,  2D $p+id$ second-order topological SCs may be hosted   in  doped  second-order topological insulators, as has been recently advocated in Ref.~\cite{BroysoloHOTSC2020}. Another available route is to make use of the proximity effect between a topological $p$-wave SC and a $d$-wave SC. Typical examples of the latter are cuprates such as YBa$_2$Cu$_3$O$_{7-\delta}$~\cite{Tsuei2000}, while the former is more difficult to find. Indeed, the leading candidate for topological $p$-wave superconductivity, Sr$_2$RuO$_4$, has recently been shown most likely \emph{not} to be of $p-$wave nature~\cite{Pustogow2019,Chronister2021}. Intensive research is, however, still ongoing and other potential candidates are the uranium-based heavy-fermion compounds, of which UTe$_2$ seems to be most promising one~\cite{Jiao2020}.

As we are concerned here only with 2D effects, we do not have to rely on bulk superconducting properties. This broadens the choice of materials somewhat. For instance, topological superconductivity in surface states can be achieved via the Fu-Kane construction~\cite{Fu-Kane2008}, whereby a conventional $s$-wave SC is placed in proximity with a topological insulator. Such a system can host edge states upon breaking of time reversal symmetry, e.g. by introducing a magnetic vortex. This behavior has recently been discovered to occur at surfaces of some materials, such as the iron-based SC FeTe$_{1-x}$Se$_x$~\cite{Xu2016, Wang2017, Zhang2018, Kong2019, Machida2019, Zhu2020}, and is anticipated in PbSaTe$_2$~\cite{Guan2021}, which thus places them as promising candidates for our purposes. Hence,  our Hamiltonian may effectively describe  the Majorana bound states on the surface of such materials, when they are proximitized with a $d$-wave SC, either as a bilayer, or as a Josephson coupling, as was suggested in Ref.~\cite{Wang2018}. We also point out that a proximity effect between niobium and graphene, through a single layer of chiral molecules, has been demonstrated to display signatures which may be indicative of chiral $p$-wave superconductivity~\cite{Sukenik2018}, and might provide additional means of realizing in the future the  model we studied here.

Finally, we remark that a cuprate $d$-wave SC typically has a significantly larger critical temperature than the  $p$-wave SCs discussed above. For instance, YBa$_2$Cu$_3$O$_{7-\delta}$ may reach a $T_c\sim100$~{\rm K}~\cite{Hussain2008}, whereas FeTe$_{0.55}$Se$_{0.45}$ has $T_c\sim15$~{\rm K}~\cite{Zhang2018}, and $T_c$ is doping-dependent in both cases. Hence, the relative size of the two superconducting order parameters, $\Delta_p$ and $\Delta_d$, might be tunable by changing the temperature and the doping.
In any case, we hope that our findings will motivate further experimental efforts to demonstrate the tunability of the  MZMs in mixed-pairing  topological SCs by both the parent $p$-wave state and an external nonthermal tuning parameter, such as strain. Our results should also motivate further searches for material platforms where the proposed scenario can be relevant.

In closing, we point out that our mechanism  may also be relevant  for three-dimensional SCs which can be realized in
doped octupolar (third-order) Dirac insulators, hosting corner
MZMs~\cite{RoyODI2021}. This problem is, however, left for future investigation.

\vspace{2ex}
\begin{acknowledgments}
M.A. thanks Jeroen Danon for useful discussions. V.J. is thankful to Bitan Roy for useful discussions and the critical reading of the manuscript.
The authors thank Henrik Roising for useful discussions.
V.J. acknowledges support
of the Swedish Research Council (VR 2019-04735).
\end{acknowledgments}



\end{document}